\documentclass[reprint,aps,prx,superscriptaddress,nofootinbib,longbibliography,floatfix]{revtex4-2}

\usepackage[T1]{fontenc}
\usepackage[utf8]{inputenc}
\usepackage{amsmath,amssymb,amsfonts,bm,mathtools}
\usepackage{graphicx}
\usepackage{xcolor}
\usepackage{microtype}
\usepackage{url}
\usepackage{CJK}
\usepackage[colorlinks=true,linkcolor=blue,citecolor=blue,urlcolor=blue]{hyperref}

\newcommand{\El}{s}
\newcommand{\Budget}{\mathcal{B}}
\newcommand{\dd}{\mathrm{d}}
\newcommand{\red}{\mathrm{red}}
\newcommand{\abs}[1]{\left|#1\right|}

\newcommand{\mfpt}{\tau}
\newcommand{\toB}{{\scriptscriptstyle A\to B}}
\newcommand{\one}{\mathbf{1}}       
\newcommand{\vect}[1]{\boldsymbol{#1}} 

\begin{document}
\begin{CJK*}{UTF8}{gbsn}
\title{A Universal Control Budget for First-Passage Kinetics}

\author{Shiling Liang (梁师翎)}
\email{shiling@pks.mpg.de}
\affiliation{Center for Systems Biology Dresden, 01307 Dresden, Germany}
\affiliation{Max Planck Institute for the Physics of Complex Systems, 01187 Dresden, Germany}
\affiliation{Max Planck Institute of Molecular Cell Biology and Genetics, 01307 Dresden, Germany}

\author{Ruicheng Bao}
\email{ruicheng@g.ecc.u-tokyo.ac.jp}
\affiliation{Department of Physics, Graduate School of Science, The University of Tokyo, 7-3-1 Hongo, Bunkyo-ku, Tokyo 113-0033, Japan}

\date{\today}

\begin{abstract}
The first-passage time is the natural observable of reaction completion, yet how its mean responds to a rate change has lacked a general constraint. We show that the logarithmic sensitivity of the mean first-passage time of any finite Markov chain to any rate is bounded by one in magnitude, and that these sensitivities sum to $-1$. Together the two laws form a conserved \emph{control budget}: speeding completion through some transitions must be paid for by others, and a coordinated change shifts the completion time only as far as the budget allows. Raising an activation barrier or shifting the depth of a well moves many rates at once, yet neither can shift the completion time further than a single rate could. The budget caps kinetic-proofreading discrimination at the checkpoint count, and prices it in sensitivity to substrate concentration.
\end{abstract}

\maketitle
\end{CJK*}
\noindent
\section{Introduction}
The first-passage time, the stochastic time for a system to first reach a target state, governs the timming for an enormous range of physical and biological processes~\cite{Redner2001,CondaminEtAl2007,BenichouVoituriez2014}. A reaction completes, an ion channel opens, a molecular motor steps, or an enzyme releases product when a hidden coordinate first crosses a threshold, so the mean first-passage time (MFPT) is what experiments record as a reaction, dwell, or turnover time~\cite{Hill1989,KussiusPopescu2009,Vale2000,GromadskiRodnina2004}. More informative than the time itself is how it \emph{responds} to a perturbation: the logarithmic sensitivity of the MFPT to a microscopic rate identifies the controlling step, turns kinetic perturbation data into mechanism, and quantifies the kinetic cost of biological function. That response, though, has lacked a universal constraint of its own.

For the response of stationary currents, such laws are now known. 
Response theory for nonequilibrium steady states bounds how a stationary observable reacts to a rate perturbation~\cite{OwenGingrichHorowitz2020,BaoLiang2025}, and has since grown into a broad theory of static and dynamical response for Markov jump processes, in which the responses are constrained by graph topology, related to one another across observables, and tied to fluctuations by nonequilibrium fluctuation--response relations~\cite{AslyamovEsposito2024a,AslyamovEsposito2024,HarunariEtAl2024,DalCengioEtAl2025,BebonSpeck2026,WangWangRen2026,FancherHorowitz2026,KatayamaNagayamaIto2026,GoerlichEtAl2026,KlettLindner2025}. Universal response inequalities now bound response by dissipation or by dynamical activity, and reach away from stationarity, to finite-frequency and to kinetic as well as entropic perturbations~\cite{AslyamovPtaszynskiEsposito2025,LiuGu2025,ZhengLu2025a,ZhengLu2025b,KwonChunParkLee2025,Dechant2026,GaoChunHorowitz2024}. All of this answers what controls a steady-state current of state observables, and by how much.

First-passage quantities have in fact already proved useful in building the response theory, since the response of a stationary observable can be written through mean first-passage times~\cite{KhodabandehlouMaesNetocny2025,BaoLiang2025}. 
In this work we run that bridge in the opposite direction. Sending every absorption event at the target back to the source turns first passage into a stationary current whose recycled value is the reciprocal MFPT~\cite{Hill1989,KampSzabo1988}, and the completion time thereby comes under the steady-state response theory sketched above. Two rules then govern how it reacts to a rate perturbation: a local rule, that the sensitivity to any one rate is bounded by one in magnitude, and a global summation rule, that all sensitivities sum to $-1$. Together these form a conserved control budget, in which speeding completion through some transitions must be paid for through others, and a coordinated change spends only as much as the budget holds. Raising an activation barrier scales both directions of one edge, and shifting the depth of a well scales every rate leaving one state; each moves many rates at once, yet neither can shift the completion time further than a single rate could. We use this budget to show that kinetic proofreading~\cite{Hopfield1974,Ninio1975,RaoPeliti2015} can discriminate by at most its checkpoint count and pays for that discrimination with a fragile dependence on substrate concentration~\cite{KumarBanerjeeGangopadhyay2022}.


\begin{figure}[!htb]
\centering
\includegraphics[width=\columnwidth]{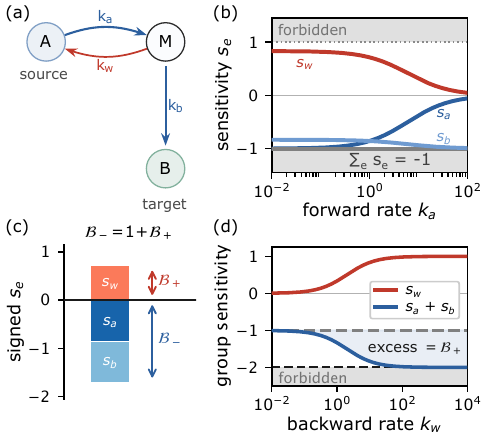}
\caption{\textbf{The two response laws and the conserved budget.}
(a) Minimal completion problem: a chain from source $A$ through an intermediate $M$ to absorbing target $B$, with forward edges (blue) and a backward edge (red).
(b) Each productive (blue) edge's sensitivity measures how rate-limiting it is: a sigmoid in the log-rate approaching the unit bound only as that edge becomes the bottleneck; the delaying (red) edge carries the complementary stored-budget. All sensitivities sum to $-1$.
(c) The conserved budget: signed sensitivities stack into a speeding budget $\Budget_-$ and a delaying budget $\Budget_+$ differing by exactly one, $\Budget_-=1+\Budget_+$.
(d) Coordinated control beats the single-edge bound: scaling both forward rates together gives $-1$ in the feedforward chain but approaches $-2$ as the backward rate grows, the excess paid by the backward edge's delaying budget.}
\label{fig:laws}
\end{figure}

\section{MFPT response laws}
Consider a continuous-time Markov jump process $\{X_t\}_{t\ge0}$ on a
finite state space, with 
a rate $k_e$ on
each directed edge $e:n_e\to m_e$. Fix a source state $A$ and a target
state $B$, and let
\begin{equation}
  \mfpt_\toB=\mathbb{E}_A\!\left[\inf\{t\ge0:X_t=B\}\right]
\end{equation}
be the mean first-passage time (MFPT) from $A$ to
$B$. Throughout this work we also refer to $\mfpt_\toB$ as the \emph{completion time}, since in every application we consider --- a reaction running to product, an enzyme releasing substrate, a receptor reaching its signaling state --- arrival at $B$ is the event that completes the process. Every sensitivity and budget below refers to this single quantity. For an irreducible Markov chain, deleting the edges out of $B$ makes it absorbing without altering
any first-passage statistic to $B$~\cite{Sekimoto2021}, so we take $B$
to be absorbing without loss of generality. 
We assume throughout that the retained class is finite with positive rates, and that $B$ is reached from every one of its states with probability one in finite mean time. The retained edges form the full directed edge set $\mathcal{C}$ of the absorbing network, over which the summation in Eq.~(3) runs; a perturbation may act on any subset of $\mathcal{C}$. Perturbing the rate of one edge $e$ then defines its \emph{sensitivity}, the logarithmic derivative
\begin{equation}
\El_e\equiv\frac{\partial\log\mfpt_\toB}{\partial\log k_e}.
\end{equation}
A negative $\El_e$ marks an edge that speeds completion when accelerated; a positive $\El_e$ marks a reset, rejection, or backward step that delays it. 
The sensitivity obeys two laws,
\begin{equation}
-1\le\El_e\le1,\quad \sum_{e\in\mathcal{C}}\El_e=-1.
\label{eq:laws}
\end{equation}
The first law is local: each edge, on its own, can change the completion time by at most its own fractional change. The second is global: the sensitivities to all rates sum to the response of a uniformly rescaled clock. 

\begin{figure}[!tbh]
\centering
\includegraphics[width=\columnwidth]{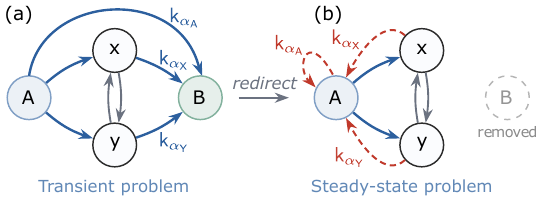}
\caption{\textbf{Redirection maps first passage to a stationary current.}
(a) Transient problem: trajectories run from the source $A$ until absorption at $B$. 
(b) Redirected graph: each absorption channel $\alpha$ is sent back to the source $A$ at the same rate and $B$ is removed, so completions recycle and the stationary rate of redirected absorption is the recycled current.}
\label{fig:redirect}
\end{figure}

The two laws have different origins. Scaling every rate by $\lambda$ rescales the clock, $\mfpt_\toB(\lambda k)=\lambda^{-1}\mfpt_\toB(k)$, so $\mfpt_\toB$ is homogeneous of degree $-1$ in the rates and Euler's theorem gives $\sum_{e\in\mathcal{C}}\El_e=-1$ immediately. The unit bound needs more, and we obtain it by mapping the absorbing chain onto an irreducible one: redirect every edge that ends at the target back to the source $A$~\cite{Hill1989} (Fig.~\ref{fig:redirect}). Let $\mathcal{A}\subset\mathcal{C}$ be the absorption edges, those ending at the target, indexed by $\alpha$ with rate $k_\alpha$ and source state $n_\alpha$. Deleting $B$ and sending each of them back to $A$ at the same rate leaves every other edge untouched. We mark all quantities of the redirected chain with $\tilde{\cdot}$ throughout. The redirected destination of an edge is $\tilde m_e=m_e$ off $\mathcal{A}$ and $\tilde m_\alpha=A$ on it, and a direct $A\to B$ edge becomes a counted self-return at $A$. The redirected network is irreducible on the retained states, so by the Perron--Frobenius theorem it has a unique stationary distribution $\tilde\pi$. Exactly one redirected event closes each completed passage, so the stationary rate of redirected absorption is a recycled current equal to the reciprocal MFPT~\cite{Hill1989},
\begin{equation}
J_\toB^{\red}=\sum_{\alpha\in\mathcal{A}}\tilde{j}_\alpha=\frac{1}{\mfpt_\toB},\quad \tilde{j}_\alpha\equiv k_\alpha\,\tilde\pi_{n_\alpha},
\label{eq:current}
\end{equation}
with $\tilde{j}_\alpha\ge0$ the one-way flux carried by channel $\alpha$, and more generally $\tilde j_e\equiv k_e\tilde\pi_{n_e}$ for any retained edge. We call $\tilde j_e$ the traffic of $e$ throughout: the one-way flux along a single directed edge. Because each renewal cycle of the redirected process is one passage from $A$ to absorption, the redirected stationary distribution is proportional to the expected occupation time before completion: $\tilde\pi_x=\nu_x/\mfpt_\toB$, where $\nu_x=\mathbb{E}_A[\int_0^{T_B}\mathbf{1}\{X_t=x\}\,\dd t]$ is the expected time spent at state $x$ before the process finishes. In particular, $\tilde j_e\,\mfpt_\toB=k_e\nu_{n_e}$ is the expected number of traversals of edge $e$ per completion.
Equation~\eqref{eq:current} collapses first passage to a single stationary current, and identifies the completion sensitivity with minus that current's response, $\El_e=-\partial\log J_\toB^{\red}/\partial\log k_e$.

The redirected network is a genuine steady state, so the response identity for Markov jump processes~\cite{BaoLiang2025} applies to each of its channel fluxes,
\begin{equation}
\frac{\partial\log\tilde{j}_\alpha}{\partial\log k_e}=\left(\tilde{\tau}_{n_e\to n_\alpha}-\tilde{\tau}_{\tilde{m}_e\to n_\alpha}\right)\tilde{j}_e+\delta_{e\alpha},
\label{eq:response}
\end{equation}
where $\tilde\tau_{x\to y}$ is the MFPT from $x$ to $y$ on the redirected graph, with $\tilde\tau_{x\to x}=0$, and the Kronecker delta $\delta_{e\alpha}$ carries the explicit rate factor an absorption channel has in its own flux. Two kinetic inequalities bound the right-hand side. Applying the MFPT triangle inequality in both orders gives $-\tilde\tau_{\tilde m_e\to n_e}\le\tilde\tau_{n_e\to n_\alpha}-\tilde\tau_{\tilde m_e\to n_\alpha}\le\tilde\tau_{n_e\to\tilde m_e}$, so the difference of hitting times is squeezed between the two legs of the perturbed edge's commute. The traffic--MFPT inequality then bounds that commute against the traffic itself, $\tilde j_e(\tilde\tau_{n_e\to\tilde m_e}+\tilde\tau_{\tilde m_e\to n_e})\le1$, because a firing leaves the process at the destination and it must return to the source before firing again, while the wait from there is at least the first-passage time back to the destination, a direct firing being itself one way of arriving : formally, the firings of $e$ form a stationary point process of intensity $\tilde j_e$, and the Kac recurrence formula~\cite{Cox1962,DaleyVereJones2003} equates intensity times mean inter-firing interval to one; the inter-firing interval decomposes into the return from $\tilde m_e$ to $n_e$ and the subsequent wait for $e$ to fire, and since a direct firing is one route from $n_e$ to $\tilde m_e$ the wait exceeds $\tilde\tau_{n_e\to\tilde m_e}$, so the commute time is at most the inter-firing interval. Together the two put the difference term in $[-1,1]$, and for $e\neq\alpha$ that is the whole response. For the perturbed channel itself $\tilde\tau_{n_e\to n_e}=0$ leaves the term equal to $-\tilde j_e\tilde\tau_{\tilde m_e\to n_e}$, which lies in $[-1,0]$, so with $\delta_{e\alpha}=1$ the response lies in $[0,1]$. Every channel-flux response is therefore in $[-1,1]$. The recycled current is a positive sum of those fluxes, so its own response is a convex combination of theirs and lies in the same interval, which is the unit bound $-1\le\El_e\le1$ of Eq.~\eqref{eq:laws}.


\section{The control budget}
Taken together the two laws fix what it costs to make a completion time respond strongly to a chosen set of rates. Sorting the edges by the sign of their sensitivity turns the pair into an accounting identity: define the speeding and delaying budgets
\begin{equation}
\Budget_+=\sum_{e: s_e>0}\El_e,\quad \Budget_-=\sum_{e:s_e<0}-\El_e,
\end{equation}
and the summation rule of Eq.~\eqref{eq:laws} becomes a conservation law,
\begin{equation}
\Budget_-=1+\Budget_+.
\label{eq:budget}
\end{equation}
Both $\Budget_+$ and $\Budget_-$ are nonnegative by construction. Speeding sensitivity exceeds delaying sensitivity by exactly one unit, the gap fixed by the summation rule (Eq.\eqref{eq:laws}). Scaling a group of edges by a common factor moves the MFPT by the sum of their sensitivities, at most the budget that group carries, so every unit of speeding beyond the first has to be bought by loading the opposite-sign edges to the same amount. Fixing this accounting requires combining both laws: the summation rule permits arbitrarily large sensitivities provided they cancel, whereas the unit bound constrains each edge individually. 

The local bound also makes the budget finite. Each edge contributes at most one, so each budget is capped by the number of edges carrying its sign, and the conservation law caps it a second time through the opposite count,
\begin{equation}
\Budget_+\le \min(N_+,\,N_- - 1),\quad \Budget_-\le \min(N_-,\,N_+ + 1),
\label{eq:edgecount}
\end{equation}
where $N_\pm$ count the edges with $\El_e\gtrless0$ at the current operating point, $N_++N_-\le\abs{\mathcal{C}}$, and every remaining edge has $\El_e=0$. These counts are themselves an operating-point property, because an edge on a cycle can speed completion at one point and delay it at another. A completion time can therefore be made strongly sensitive to a coordinated perturbation only by spreading the response over many edges. Metabolic control analysis gives the analogous summation theorem~\cite{KacserBurns1973,HeinrichRapoport1974,Fell1992,MelendezHeviaEtAl1990,Liu2025MCA} but has no local unit bound, so it sets no scale for an individual control and yields no budget.

The smallest completion problem shows the budget being spent (Fig.~\ref{fig:laws}). The feedforward chain $A\to M\to B$, with forward rates $k_a,k_b$ and $B$ absorbing, has $\mfpt_\toB=1/k_a+1/k_b$. Nothing in it delays completion, so $\Budget_+=0$ and Eq.~\eqref{eq:budget} leaves a single unit of speeding budget for the two forward edges to share. Adding a backward step $M\to A$ at rate $k_w$ opens a delaying budget and gives $\mfpt_\toB=(k_a+k_w+k_b)/(k_a k_b)$. As $k_w$ grows the trajectory idles in the $A\rightleftharpoons M$ cycle before escaping through $M\to B$, and the backward edge fills its budget, $\El_w\to+1$. Conservation then gives $\mathcal{B}_{-}=-s_a-s_b\to2$. Thus, scaling only $k_a$ and $k_b$ together while holding $k_w$ fixed yields the group sensitivity $s_a+s_b=-\mathcal{B}_{-}$, which approaches $-2$ as $k_w\to\infty$: twice what any single edge can achieve, with the extra unit supplied by the delaying budget stored in the backward edge (Fig.~\ref{fig:laws}d).


\section{Barrier and state controls}
So far we have considered perturbations of individual edge rates and arbitrary groups of them. Many physical controls instead change a specific group of rates fixed by the underlying mechanism. Catalysis or a mutation at a transition state can shift a kinetic barrier and thereby scale the two opposite rates across one link together, whereas stabilizing a conformation or a bound complex can shift a state energy and thereby scale every rate leaving that state. We define a \emph{barrier perturbation} by $(k_{u\to v},k_{v\to u})\mapsto e^\theta(k_{u\to v},k_{v\to u})$, with response $\El^{\leftrightarrow}_{uv}\equiv\left.\partial_\theta\log\mfpt_\toB\right|_{\theta=0}$. We define a \emph{state perturbation} at $v$ by $k_e\mapsto e^\theta k_e$ for every edge $e$ with $n_e=v$, with response $\El^{\mathrm{out}}_v\equiv\left.\partial_\theta\log\mfpt_\toB\right|_{\theta=0}$. Both responses are bounded by one in magnitude, but for different reasons: cancellation between the two directions limits a barrier response, whereas the magnitude of a state response is exactly that state's share the completion time.

Both results follow from a rewriting of Eq.~\eqref{eq:response} in the original absorbing chain. {The channel-averaged redirected hitting time reproduces the remaining wait of the original chain, $\sum_\alpha\omega_\alpha(\tilde\tau_{x\to n_\alpha}-\tilde\tau_{y\to n_\alpha})=\mfpt_{x\to B}-\mfpt_{y\to B}$, with weights $\omega_\alpha=\tilde j_\alpha/J_\toB^{\red}$ (Appendix~\ref{app:channelaverage}). Averaging Eq.~\eqref{eq:response} over these channels then} gives
\begin{equation}
\El_e=\tilde j_e\left(\mfpt_{m_e\to B}-\mfpt_{n_e\to B}\right),
\label{eq:edgeclosed}
\end{equation}
where $\mfpt_{x\to B}$ is the MFPT from $x$ in the original absorbing chain and $\mfpt_{B\to B}=0$. The two structured responses defined above are the corresponding sums of the single-rate sensitivities,
$\El^{\leftrightarrow}_{uv}=\El_{u\to v}+\El_{v\to u}$ and $\El^{\mathrm{out}}_v=\sum_{e:\,n_e=v}\El_e$. 

 Since $\tilde j_e\,\mfpt_\toB$ is the expected number of traversals of $e$ per completion (Sec.~II), Eq.~\eqref{eq:edgeclosed} says that an edge controls the completion time in proportion to how often the trajectory uses it and how much remaining wait each traversal removes. Summed over all edges, the differences $\mfpt_{m_e\to B}-\mfpt_{n_e\to B}$ telescope from $\mfpt_\toB$ to zero along every trajectory, which returns the summation rule as an exact signed decomposition of the completion time.

\begin{figure}[!htb]
\centering
\includegraphics[width=\columnwidth]{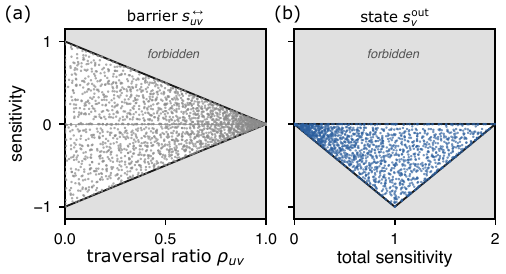}
\caption{\textbf{Barrier and state controls keep the unit bound.} Sensitivity of one multi-rate control per panel, on a common ordinate, for randomly generated completion networks. Shaded regions are forbidden.
(a) Barrier control $\El^{\leftrightarrow}_{uv}$ against the traversal ratio $\rho_{uv}$ with the envelope of Eq.~\eqref{eq:barrier}.
(b) State control $\El^{\mathrm{out}}_v$ against the total sensitivity $\sum_{e:\,n_e=v}\abs{\El_e}$ of the rates it moves, taken before any cancellation of sign. The bounding lines are Eq.~\eqref{eq:statecap} and $\sum_{e:\,n_e=v}\abs{\El_e}\ge\abs{\El^{\mathrm{out}}_v}$.}
\label{fig:controls}
\end{figure}

For the barrier perturbation defined above, the two directions carry opposite MFPT differences, so by Eq.~\eqref{eq:edgeclosed} their sensitivities have opposite signs and only the \emph{net} current survives, $\El^{\leftrightarrow}_{uv}=(\tilde j_{u\to v}-\tilde j_{v\to u})(\mfpt_{v\to B}-\mfpt_{u\to B})$. Writing $\rho_{uv} = \min\left( \frac{\tilde{j}_{u\to v}}{\tilde{j}_{v\to u}}, \, \frac{\tilde{j}_{v\to u}}{\tilde{j}_{u\to v}} \right) \in [0, 1]$
for the \emph{traversal ratio} of the edge, the smaller directed traffic over the larger and equivalently the ratio of the expected traversals in the two directions, the two single-rate bounds give
\begin{equation}
\big|\El^{\leftrightarrow}_{uv}\big|=(1-\rho_{uv})\max\!\left(\abs{\El_{u\to v}},\abs{\El_{v\to u}}\right)\le1-\rho_{uv}.
\label{eq:barrier}
\end{equation}
A barrier controls completion only in proportion to the net current across it. An edge the trajectory shuttles across equally in both directions is invisible to its own barrier, however rate-limiting each of its two rates is separately: the chain of Fig.~\ref{fig:laws}d shows this sharply, because in the limit $k_w\to\infty$, precisely where every forward rate reaches $\El_e\to-1$, every barrier response vanishes. Nor is the sign fixed: when the net current runs away from the target, raising the barrier speeds completion by suppressing a futile excursion. It requires no thermodynamic driving and happens already on a symmetric edge, since absorption alone is enough to 
yield a nonzero net current $\tilde j_{u\to v}-\tilde j_{v\to u}$. Figure~\ref{fig:controls}a shows the envelope approached across random networks.

For the state perturbation defined above, summing Eq.~\eqref{eq:edgeclosed} over the edges leaving $v$ and using the backward equation $\sum_{e:\,n_e=v}k_e(\mfpt_{m_e\to B}-\mfpt_{v\to B})=-1$ collapses the group response to an identity:
\begin{equation}
\El^{\mathrm{out}}_v=\sum_{e:\,n_e=v}\El_e=-\tilde\pi_v\in[-1,0],\quad \sum_{v}\El^{\mathrm{out}}_v=-1.
\label{eq:state}
\end{equation}
State sensitivities are therefore 
simply a partition of one: each state's share of the control is its share of the completion time, and destabilizing a state can only speed completion.

Equation~\eqref{eq:state} fixes only the signed sum of the sensitivities of the rates leaving $v$. By itself, it would still allow a large speeding contribution to be cancelled by an equally large delaying contribution. A stronger constraint follows because all these rates share the same source state. Let $\mathcal{G}$ be any subset of the edges leaving $v$, and scale all rates in $\mathcal{G}$ together. In the redirected process, $\mathcal{G}$ can be treated as a single grouped channel: it fires from $v$ with total traffic $\tilde j_{\mathcal{G}}=\sum_{e\in\mathcal{G}}\tilde j_e$ and, conditional on firing, lands at $\tilde m_e$ with probability $\tilde j_e/\tilde j_{\mathcal{G}}$. Summing Eq.~\eqref{eq:response} over $\mathcal{G}$ and applying the same triangle and traffic--MFPT inequalities used for a single edge, now averaged over these possible destinations, gives {(Appendix~\ref{app:samesource})}
\begin{equation}
\Big|\sum_{e\in\mathcal{G}}\El_e\Big|\le1,
\label{eq:samesource}
\end{equation}
regardless of how many rates $\mathcal{G}$ contains. Taking $\mathcal{G}$ to be the speeding exits alone caps the budget they carry at $\Budget^{(v)}_-\le1$, with $\Budget^{(v)}_\pm$ the budgets of Eq.~\eqref{eq:budget} restricted to the edges leaving $v$: the global budget grows without bound with network size, but no single state holds more than one unit of it. Since Eq.~\eqref{eq:state} fixes the gap between the two halves at $\Budget^{(v)}_--\Budget^{(v)}_+=\tilde\pi_v$, combining it with $\Budget^{(v)}_-\le1$ bounds the \emph{total sensitivity} of those exits, what they carry before any cancellation between the two signs,
\begin{equation}
\sum_{e:\,n_e=v}\abs{\El_e}=\Budget^{(v)}_-+\Budget^{(v)}_+\le2-\abs{\El^{\mathrm{out}}_v}.
\label{eq:statecap}
\end{equation}
This is the right edge of the accessible region in Fig.~\ref{fig:controls}b. Its left edge is the elementary statement that a signed sum cannot exceed an unsigned one, $\sum_{e:\,n_e=v}\abs{\El_e}\ge\abs{\El^{\mathrm{out}}_v}$, and together they confine every state control to a triangle that the sampled networks populate densely.

Equation~\eqref{eq:state} is one case of a statement about the whole landscape. Scaling every rate out of $v$ by $e^{\lambda_v\theta}$, for an arbitrary state function $\lambda$, shifts each well depth $E_v\mapsto E_v+\lambda_v\theta$ and summing Eq.~\eqref{eq:state} over states gives $\partial_\theta\log\mfpt_\toB|_{\theta=0}=-\sum_{v'}\lambda_{v'}\tilde\pi_{v'}$. Reshaping the landscape at fixed barriers therefore shifts the completion time by the occupancy-weighted average of the state shifts, bounded in magnitude by the largest single shift however many states move. The family interpolates between the two laws already in hand: $\lambda\equiv1$ is the uniform rescaling of all rates and $\lambda=\delta_{vv'}$ is Eq.~\eqref{eq:state}, so the summation rule and the state law are the same identity.
\begin{figure}[!htb]
\centering
\includegraphics[width=1\columnwidth]{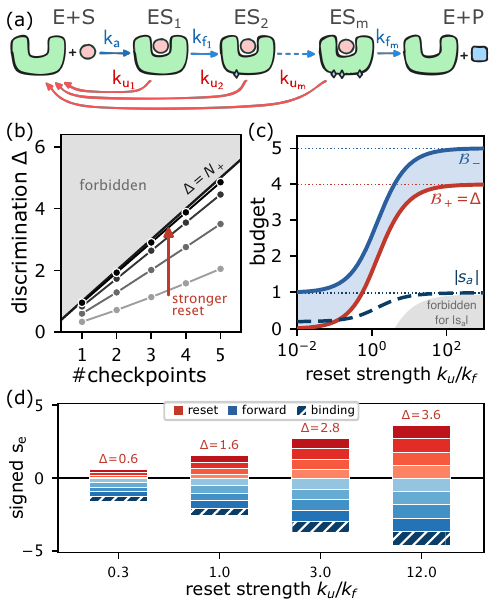}
\caption{\textbf{Proofreading spends the budget.}
(a) Binding--reset chain; the wrong substrate has faster resets, $k_{u_i}^{W}=f\,k_{u_i}^{R}$.
(b) Discrimination capacity is capped by the checkpoint count, $\Delta\le N_+=m$ (grey forbidden). The four curves are reset strengths $k_u/k_f=1,3,10,40$, light to dark, the darkest running just under the cap.
(c) The cost identity at $m=4$: the speeding budget $\Budget_-$ and the delaying budget $\Budget_+=\Delta$ rise together with a fixed unit gap $\Budget_-=1+\Delta$. The binding-edge sensitivity $|\El_a|$, the dependence on substrate concentration, is squeezed toward its unit ceiling as $\Delta$ approaches the cap; the grey region beneath it is what Eq.~\eqref{eq:fragility} excludes along this sweep.
(d) The budget bar across reset strengths at $m=4$: reset edges stack to $\Budget_+=\Delta$, the binding and forward edges to $\Budget_-=1+\Delta$, the hatched binding share approaching one as discrimination grows.
All panels use $k_a=k_{f_i}=1$ with a common reset ratio $k_{u_i}/k_{f_i}=k_u/k_f$.}
\label{fig:proof}
\end{figure}

\section{Application: Kinetic proofreading}
For a system that discriminates by first-passage timing, the budget both caps selectivity and sets its price. We make this exact in the canonical reset model of kinetic proofreading~\cite{Hopfield1974,Ninio1975,MuruganEtAl2012}, the first-passage discrimination underlying T-cell antigen recognition~\cite{McKeithan1995}. An enzyme (or receptor) binds a substrate (rate $k_a$) and advances through $m$ enzyme--substrate complexes $E_1S,\dots,E_mS$ along the forward rates $k_{f_1},\dots,k_{f_m}$, the last of which, $k_{f_m}=k_{\mathrm{cat}}$, releases product. Each complex can reset to the free enzyme (rate $k_{u_i}$), giving repeated chances for rejection (Fig.~\ref{fig:proof}a). Solving the backward equation of the kinetics recursively gives the MFPT from the initial unbound state $E+S$ to the absorbing output $E+P$:
\begin{equation}
\tau
=
\frac{1}{k_a}\prod_{i=1}^{m}\left(1+\frac{k_{u_i}}{k_{f_i}}\right)
+
\sum_{i=1}^{m}
\frac{1}{k_{f_i}}
\prod_{\ell=i+1}^{m}\left(1+\frac{k_{u_\ell}}{k_{f_\ell}}\right).
\label{eq:proofreading_tau}
\end{equation}
In the energetic-discrimination limit a correct (R) and an incorrect (W) substrate differ only in binding affinity, and tighter binding lowers the reset rates~\cite{SartoriPigolotti2013,RaoPeliti2015}. Thus, in this case, the only kinetic parameters that depend on whether the substrate is right or wrong are the reset rates, while all forward rates are identical for the two substrates. We therefore set $k_{u_i}^{W}=f\,k_{u_i}^{R}$ with $f\ge1$, and write the time-discrimination ratio using the completion time for two different substrates, $D=\mfpt^{W}/\mfpt^{R}$, which measures how much longer the wrong substrate takes to complete the process than the right one. At $f=1$ the two substrates are kinetically identical and $D=1$; as $f$ grows the wrong substrate resets more often and $D$ increases.

We define the \emph{discrimination gain} \cite{MuruganEtAl2012} $\Delta\equiv\partial\log D/\partial\log f|_{f=1}$: it measures how effectively the network amplifies a small kinetic difference between the two substrates into a difference in completion times. For instance, $\Delta=3$ means that a $1\%$ increase in the wrong substrate's reset rates produces a $3\%$ increase in $D$. Because only the reset rates depend on substrate identity, $\Delta$ is exactly the sum of the reset-edge sensitivities, $\Delta=\sum_{i=1}^{m}\El_{u_i}$. The reset edges are the positive-sensitivity edges of the correct-substrate completion time, since every productive rate $k_a,k_{f_i}$ enters Eq.~\eqref{eq:proofreading_tau} only through positive inverse factors and so has negative sensitivity, and they number one per checkpoint. The discrimination gain is therefore the delaying budget itself, $\Delta=\Budget_+$, and the edge-count bound of Eq.~\eqref{eq:edgecount} with $N_+=m$ caps it at the number of checkpoints:
\begin{equation}
\Delta\equiv\left.\frac{\partial\log D}{\partial\log f}\right|_{f=1}=\sum_{i=1}^{m}\El_{u_i}=\Budget_+\le m.
\label{eq:discrimination}
\end{equation}
The finite form is $D\le f^{m}$, one power of selectivity per checkpoint. The cap binds whatever the rates: Fig.~\ref{fig:proof}b tracks $\Delta$ against the checkpoint count as the resets strengthen, and the strongest designs run just under the bound. When substrate identity also perturbs forward or catalytic rates, namely kinetic discrimination also plays a role, the discrimination gain becomes the directional projection $\Delta_g=\sum_e g_e\El_e$, bounded by $\sum_e\abs{g_e}$ due to the unit bound on individual $s_e$; the identity $\Delta=\Budget_+\le m$ is the energetic-discrimination limit, $g$ supported on the reset edges.

The discrimination gain is bounded above, but it is not free to approach that bound without consequence. The conservation law of Eq.~\eqref{eq:budget} ties the two halves of the budget together: raising the delaying budget to $\Delta$ forces the speeding budget to $\Budget_-=1+\Delta$. In this network the speeding budget is carried entirely by the binding edge and the $m$ forward edges, since these are the only negative-sensitivity edges. The local bound of Eq.~\eqref{eq:laws} lets each forward edge absorb at most one unit, so the binding edge must carry the remainder:
\begin{equation}
\abs{\El_a}=\Budget_--\sum_{i=1}^{m}\abs{\El_{f_i}}\ \ge\ \max\!\left(0,\,1+\Delta-m\right).
\label{eq:fragility}
\end{equation}
This remainder is the sensitivity of the correct-substrate completion time to substrate concentration. Binding is set by concentration through $k_a=k_{\mathrm{on}}[S]$, so $\El_a=\partial\log\mfpt^{R}/\partial\log[S]$. Strong discrimination requires large reset rates, which repeatedly return the correct substrate to the free enzyme and force it to rebind; the completion time then comes to be dominated by the time spent waiting to rebind, which is controlled by concentration alone. Equation~\eqref{eq:fragility} makes the tradeoff sharp. A network operating within $\varepsilon$ of the checkpoint cap, $\Delta=m-\varepsilon$, is forced to $\abs{\El_a}\ge1-\varepsilon$: nearly the entire completion time is spent waiting at the binding step. In the saturating limit $\Delta\to m$ this bound and the unit bound close on $\abs{\El_a}\to1$, so the completion time falls in inverse proportion to substrate concentration, $\mfpt^{R}\propto[S]^{-1}$. Discrimination and concentration robustness are therefore not independent design choices: a network tuned for near-maximal selectivity cannot also be insensitive to the concentration of the substrate it is discriminating.

\section{Discussion and Outlook}
Redirection maps the completion time to a stationary current, allowing steady-state response bounds to be applied to MFPT sensitivities. It recasts every local timing assay as a readout of where the budget currently sits, since a kinetic-perturbation experiment measures not a rate in isolation but the share of the completion time that rate controls, and the conservation law then fixes what the remaining steps must carry. Any completion process with a resolved kinetic scheme is open to the same accounting. Target search and cleavage by CRISPR-Cas9~\cite{EslamiMossallam2022}, stepping by processive motors~\cite{Vale2000,ClancyEtAl2011,Rief2000}, activation of ligand-gated ion channels~\cite{KussiusPopescu2009}, and tRNA selection on the ribosome~\cite{GromadskiRodnina2004} all have fitted rate models, and all are probed by the single-rate perturbations the budget interprets. When a control parameter moves the operating point, as substrate or ligand concentration does through a binding edge, the budget predicts that the controlling share relocates across the network rather than appearing or vanishing. The accounting is therefore at once a diagnostic, locating the rate-limiting step and tracking where it goes as conditions change, and a design constraint, capping how sensitively any completion time can be tuned and pricing selectivity in the fragility it imposes.

The homogeneity behind the summation rule is not particular to the mean. Higher completion moments, the precision, splitting probabilities and conditional completion times each carry their own scale weight, so each invites a summation rule of its own. Whether the local bound travels with them is the harder question. First-passage times already carry thermodynamic structure of their own, from uncertainty relations~\cite{GingrichHorowitz2017,PalReuveniRahav2021,HiuraSasa2021} to an exact fluctuation symmetry around a cycle~\cite{BusielloLiangPigolotti2026}; what dissipation does to their \emph{response}, and whether a sharpened bound reaches full first-passage distributions and semi-Markov dynamics, is the natural next step.

\begin{acknowledgments}
S.~L. acknowledges financial support from the Max Planck Society. R.~B. was supported by JSPS KAKENHI Grant No. 25KJ0766. We thank Claude Opus 5.6 and GPT 5.6 for assistance with coding and editing the manuscript. 
\end{acknowledgments}

\appendix
\section{Channel-average identity}
\label{app:channelaverage}

Remove the absorbing target $B$ and let $Q$ denote the generator restricted to the retained transient states, with column sums $\sum_{y}Q_{yx}=-\kappa_x$, where $\kappa_x=\sum_{\alpha\in\mathcal{A}:\,n_\alpha=x}k_\alpha$ is the total absorption rate from state~$x$.  Collect these rates into~$\vect\kappa$ and write $\vect\delta_A$ for the coordinate vector on~$A$.  The remaining waits $\mfpt_{x\to B}$, collected into~$\vect\mfpt$, satisfy
\begin{equation}
  \vect\mfpt^{\,\top}Q=-\one^\top, \qquad \mfpt_{B\to B}=0,
  \label{eq:app:backward}
\end{equation}
with $\vect\mfpt^{\,\top}\vect\delta_A=\mfpt_\toB$.  The redirected generator
\begin{equation}
  \widetilde W=Q+\vect\delta_A\vect\kappa^{\,\top}
  \label{eq:app:Wtilde}
\end{equation}
sends every absorption jump back to $A$, making the retained chain irreducible with stationary distribution~$\tilde\pi$.

\emph{Claim.}\; The channel-averaged hitting time $\Phi_x=\sum_\alpha\omega_\alpha\,\tilde\tau_{x\to n_\alpha}$ reproduces the remaining waits: $\Phi_x-\Phi_y=\mfpt_{x\to B}-\mfpt_{y\to B}$ for all retained $x,y$.

\emph{Proof.}\; Set $f_x=\Phi_x-\mfpt_{x\to B}$; it suffices to show $\vect f^{\,\top}\widetilde W=0$, since the irreducible generator~$\widetilde W$ has left null space~$\one^\top$.  From Eqs.~\eqref{eq:app:backward} and~\eqref{eq:app:Wtilde},
\begin{equation}
  \vect\mfpt^{\,\top}\widetilde W=-\one^\top+\mfpt_\toB\,\vect\kappa^{\,\top}.
\end{equation}
For the redirected chain, $\sum_z\widetilde W_{zx}\tilde\tau_{z\to y}=-1$ when $x\neq y$ (backward equation with $\tilde\tau_{y\to y}=0$); at $x=y$ the Kac recurrence formula~\cite{Cox1962,DaleyVereJones2003} for departures from~$y$ gives $\tilde\pi_y(1+\sum_z\widetilde W_{zy}\tilde\tau_{z\to y})=1$, i.e.\ $\sum_z\widetilde W_{zy}\tilde\tau_{z\to y}=1/\tilde\pi_y-1$.  Multiplying by $\omega_\alpha$ and summing over channels,
\begin{equation}
  (\vect\Phi^{\,\top}\widetilde W)_x=-1+\frac{1}{\tilde\pi_x}\!\sum_{\alpha:\,n_\alpha=x}\!\omega_\alpha=-1+\mfpt_\toB\,\kappa_x,
\end{equation}
the last step using $\omega_\alpha=k_\alpha\tilde\pi_{n_\alpha}\mfpt_\toB$ and $\sum_{\alpha:\,n_\alpha=x}k_\alpha=\kappa_x$.  The two expressions coincide, giving $\vect f^{\,\top}\widetilde W=0$.~$\square$

Averaging Eq.~\eqref{eq:response} over channels with weights $\omega_\alpha$ replaces each redirected hitting time by its channel average~$\Phi$.  By the claim, $\Phi_{n_e}-\Phi_{\tilde m_e}=\mfpt_{n_e\to B}-\mfpt_{\tilde m_e\to B}$, yielding Eq.~\eqref{eq:edgeclosed} for non-absorption edges directly and for absorption edges after the Kronecker and MFPT terms cancel at $\mfpt_{B\to B}=0$.

\section{Same-source bound}
\label{app:samesource}

Fix a retained state $v$ and a nonempty set $\mathcal{G}$ of edges leaving~$v$ on the redirected graph, with total rate $K_{\mathcal{G}}=\sum_{e\in\mathcal{G}}k_e$ and traffic $\tilde j_{\mathcal{G}}=\tilde\pi_v K_{\mathcal{G}}$.  The average post-jump hitting time is
\begin{equation}
  \tilde\tau_{\mathcal{G}\to x}=\sum_{e\in\mathcal{G}}\frac{k_e}{K_{\mathcal{G}}}\,\tilde\tau_{\tilde m_e\to x},
  \label{eq:app:grouplanding}
\end{equation}
and $\tilde\tau_{v\to\mathcal{G}}$ denotes the mean time from~$v$ until the first firing of any edge in~$\mathcal{G}$.

\emph{Lemma (group Kac identity).}\; $\tilde j_{\mathcal{G}}\bigl(\tilde\tau_{v\to\mathcal{G}}+\tilde\tau_{\mathcal{G}\to v}\bigr)=1$.

\emph{Proof.}\; The firings of $\mathcal{G}$ form a stationary point process of intensity $\tilde j_{\mathcal{G}}$.  The Palm inversion formula~\cite{Cox1962,DaleyVereJones2003} gives $\tilde j_{\mathcal{G}}\,\mathbb{E}^0 S=1$, where $S$ is the inter-firing interval under the Palm distribution.  Since every edge of $\mathcal{G}$ leaves~$v$, the interval decomposes as $S=T_v+R$: a return to~$v$ (mean $\tilde\tau_{\mathcal{G}\to v}$) followed by a wait for the group to fire (mean $\tilde\tau_{v\to\mathcal{G}}$, by the strong Markov property).~$\square$

The Kac identity splits one unit between the two terms.  Define the \emph{waiting share} $\phi_{\mathcal{G}}=\tilde j_{\mathcal{G}}\,\tilde\tau_{v\to\mathcal{G}}\in[0,1]$, so that $1-\phi_{\mathcal{G}}=\tilde j_{\mathcal{G}}\,\tilde\tau_{\mathcal{G}\to v}$.  Standard MFPT triangle inequalities give, for any state~$x$,
\begin{equation}
  -\tilde\tau_{\mathcal{G}\to v}
  \;\le\;
  \tilde\tau_{v\to x}-\tilde\tau_{\mathcal{G}\to x}
  \;\le\;
  \tilde\tau_{v\to\mathcal{G}}.
  \label{eq:app:grouptriangle}
\end{equation}
The right bound holds because one route from $v$ to $x$ passes through a firing of~$\mathcal{G}$, giving $\tilde\tau_{v\to x}\le\tilde\tau_{v\to\mathcal{G}}+\tilde\tau_{\mathcal{G}\to x}$; the left bound follows from $\tilde\tau_{\tilde m_e\to x}\le\tilde\tau_{\tilde m_e\to v}+\tilde\tau_{v\to x}$ averaged over edges.  Multiplying by $\tilde j_{\mathcal{G}}$ and using the waiting share confines the product to a unit-width window:
\begin{equation}
  \phi_{\mathcal{G}}-1
  \;\le\;
  \bigl(\tilde\tau_{v\to x}-\tilde\tau_{\mathcal{G}\to x}\bigr)\tilde j_{\mathcal{G}}
  \;\le\;
  \phi_{\mathcal{G}}.
  \label{eq:app:groupwindow}
\end{equation}

\emph{Theorem.}\; $\bigl\lvert\sum_{e\in\mathcal{G}}\El_e\bigr\rvert\le1$.

\emph{Proof.}\; Summing Eq.~\eqref{eq:response} over $e\in\mathcal{G}$ (all sharing $n_e=v$) gives
\begin{equation}
  \sum_{e\in\mathcal{G}}\frac{\partial\log\tilde j_\alpha}{\partial\log k_e}
  =\bigl(\tilde\tau_{v\to n_\alpha}-\tilde\tau_{\mathcal{G}\to n_\alpha}\bigr)\tilde j_{\mathcal{G}}
  +\sum_{e\in\mathcal{G}}\delta_{e\alpha}.
  \label{eq:app:groupresponse}
\end{equation}
For $\alpha\notin\mathcal{G}$ the Kronecker sum vanishes and the response lies in $[\phi_{\mathcal{G}}-1,\,\phi_{\mathcal{G}}]$ by Eq.~\eqref{eq:app:groupwindow}.  For $\alpha\in\mathcal{G}$ we have $n_\alpha=v$ and $\tilde\tau_{v\to v}=0$, so the difference term equals $\phi_{\mathcal{G}}-1$ and the Kronecker delta shifts it to~$\phi_{\mathcal{G}}$.  Every channel therefore lies in the unit-width window $[\phi_{\mathcal{G}}-1,\,\phi_{\mathcal{G}}]$.  Since $\sum_{e\in\mathcal{G}}\El_e$ is minus the convex combination $\sum_\alpha\omega_\alpha(\cdots)$ with $\omega_\alpha\ge0$, $\sum_\alpha\omega_\alpha=1$, the result lies in $[-\phi_{\mathcal{G}},\,1-\phi_{\mathcal{G}}]\subseteq[-1,1]$.~$\square$

\section{Proofreading closed forms}
\label{app:proofreading}

Split the completion time of Eq.~\eqref{eq:proofreading_tau} into clock terms $\sigma_0=k_a^{-1}\prod_{\ell=1}^{m}(1+k_{u_\ell}/k_{f_\ell})$ and $\sigma_i=k_{f_i}^{-1}\prod_{\ell=i+1}^{m}(1+k_{u_\ell}/k_{f_\ell})$ for $1\le i\le m$, with prefix sums $\Sigma_i=\sum_{\ell=0}^{i-1}\sigma_\ell$.  The factor $1+k_{u_i}/k_{f_i}$ appears exactly in $\sigma_0,\ldots,\sigma_{i-1}$, so
\begin{equation}
\begin{aligned}
  &\El_{u_i}=\frac{k_{u_i}}{k_{f_i}+k_{u_i}}\frac{\Sigma_i}{\mfpt}\ge0, \quad 
  \El_a=-\frac{\sigma_0}{\mfpt}\le0,\\
  &\El_{f_i}=-\frac{1}{\mfpt}\Bigl(\frac{k_{u_i}}{k_{f_i}+k_{u_i}}\Sigma_i+\sigma_i\Bigr)\le0.
\end{aligned}
  \label{eq:app:Eclosed}
\end{equation}
The discrimination gain $\Delta=\sum_i\El_{u_i}=\Budget_+$ and the cost identity $\Budget_-=\abs{\El_a}+\sum_i\abs{\El_{f_i}}=1+\Delta$ follow by direct summation.  The concentration sensitivity is $\El_a=-\sigma_0/\mfpt$, since $k_a=k_{\mathrm{on}}[S]$.  The finite form $D\le f^m$ follows from integrating the unit bound along each single-rate ray and applying the resulting envelope $\min(c,c^{-1})\le\mfpt(ck_e)/\mfpt(k)\le\max(c,c^{-1})$ to each of the $m$ reset rates in sequence.

\section{Numerical provenance}
\label{app:numerics}

All sensitivities are computed by exact linear algebra: the occupation vector solves $Q\vect\nu=-\vect\delta_A$ with $\mfpt_\toB=\one^\top\vect\nu$, and differentiating gives $Q\,\partial\vect\nu/\partial k_e=-(\partial Q/\partial k_e)\vect\nu$, whence $\El_e=(k_e/\mfpt_\toB)\one^\top\partial\vect\nu/\partial k_e$.  Every sensitivity is cross-validated against Eq.~\eqref{eq:edgeclosed}.  Networks behind Fig.~\ref{fig:controls} use $8$--$20$ transient states with random spanning trees, extra reversible edges, and one to three absorption edges, all rates log-uniform over $10^{-4}$--$10^{4}$; instances with $\mathrm{cond}(Q)>10^{10}$ are discarded.  Figure~\ref{fig:proof} uses the closed forms of Eq.~\eqref{eq:app:Eclosed} on the homogeneous design $k_a=k_{f_i}=1$ with common reset ratio $k_{u_i}/k_{f_i}=k_u/k_f$.

\bibliography{refs}

@book{Redner2001,
  author = {S. Redner},
  title = {A Guide to First-Passage Processes},
  publisher = {Cambridge University Press},
  year = {2001},
  doi = {10.1017/CBO9780511606014}
}

@book{Hill1989,
  author    = {Hill, Terrell L.},
  title     = {Free Energy Transduction and Biochemical Cycle Kinetics},
  publisher = {Springer-Verlag},
  address   = {New York},
  year      = {1989},
  doi       = {10.1007/978-1-4612-3558-3},
  isbn      = {978-0-387-96836-0}
}

@article{CondaminEtAl2007,
  author = {S. Condamin and O. B{\'e}nichou and V. Tejedor and R. Voituriez and J. Klafter},
  title = {First-passage times in complex scale-invariant media},
  journal = {Nature},
  volume = {450},
  pages = {77--80},
  year = {2007},
  doi = {10.1038/nature06201}
}

@article{BenichouVoituriez2014,
  author = {O. B{\'e}nichou and R. Voituriez},
  title = {From first-passage times of random walks in confinement to geometry-controlled kinetics},
  journal = {Phys. Rep.},
  volume = {539},
  pages = {225--284},
  year = {2014},
  doi = {10.1016/j.physrep.2014.02.003}
}

@article{Hopfield1974,
  author = {J. J. Hopfield},
  title = {Kinetic proofreading: A new mechanism for reducing errors in biosynthetic processes requiring high specificity},
  journal = {Proc. Natl. Acad. Sci. U.S.A.},
  volume = {71},
  pages = {4135--4139},
  year = {1974},
  doi = {10.1073/pnas.71.10.4135}
}

@article{MuruganEtAl2012,
  author = {A. Murugan and D. A. Huse and S. Leibler},
  title = {Speed, dissipation, and error in kinetic proofreading},
  journal = {Proc. Natl. Acad. Sci. U.S.A.},
  volume = {109},
  pages = {12034--12039},
  year = {2012},
  doi = {10.1073/pnas.1119911109}
}

@misc{BaoLiang2025,
  author       = {Bao, Ruicheng and Liang, Shiling},
  title        = {Nonlinear response identities and bounds for nonequilibrium steady states},
  year         = {2024},
  eprint       = {2412.19602},
  archivePrefix= {arXiv},
  primaryClass = {cond-mat.stat-mech},
  url          = {https://arxiv.org/abs/2412.19602},
}

@book{Cox1962,
  author = {D. R. Cox},
  title = {Renewal Theory},
  publisher = {Methuen},
  year = {1962}
}

@book{DaleyVereJones2003,
  author = {D. J. Daley and D. Vere-Jones},
  title = {An Introduction to the Theory of Point Processes, Vol. I: Elementary Theory and Methods},
  edition = {2nd},
  publisher = {Springer},
  address = {New York},
  year = {2003},
  doi = {10.1007/b97277}
}

@article{Ninio1975,
  author = {J. Ninio},
  title = {Kinetic amplification of enzyme discrimination},
  journal = {Biochimie},
  volume = {57},
  pages = {587--595},
  year = {1975},
  doi = {10.1016/S0300-9084(75)80139-8}
}

@article{KussiusPopescu2009,
  author = {C. L. Kussius and G. K. Popescu},
  title = {Kinetic basis of partial agonism at NMDA receptors},
  journal = {Nat. Neurosci.},
  volume = {12},
  pages = {1114--1120},
  year = {2009},
  doi = {10.1038/nn.2361}
}

@article{Vale2000,
  author = {R. D. Vale and R. A. Milligan},
  title = {The way things move: Looking under the hood of molecular motor proteins},
  journal = {Science},
  volume = {288},
  pages = {88--95},
  year = {2000},
  doi = {10.1126/science.288.5463.88}
}

@article{ClancyEtAl2011,
  author = {B. E. Clancy and W. M. Behnke-Parks and J. O. L. Andreasson and S. S. Rosenfeld and S. M. Block},
  title = {A universal pathway for kinesin stepping},
  journal = {Nat. Struct. Mol. Biol.},
  volume = {18},
  pages = {1020--1027},
  year = {2011},
  doi = {10.1038/nsmb.2104}
}

@article{Rief2000,
  author = {M. Rief and R. S. Rock and A. D. Mehta and M. S. Mooseker and R. E. Cheney and J. A. Spudich},
  title = {Myosin-V stepping kinetics: A molecular model for processivity},
  journal = {Proc. Natl. Acad. Sci. U.S.A.},
  volume = {97},
  pages = {9482--9486},
  year = {2000},
  doi = {10.1073/pnas.97.17.9482}
}

@article{KampSzabo1988,
  author = {F. Kamp and A. Szabo},
  title = {Fluxes, first passage times, and the reduction of {Hill} diagrams},
  journal = {Cell Biophys.},
  volume = {12},
  pages = {145--155},
  year = {1988},
  doi = {10.1007/BF02918356}
}

@article{PalReuveniRahav2021,
  author = {A. Pal and S. Reuveni and S. Rahav},
  title = {Thermodynamic uncertainty relation for first-passage times on Markov chains},
  journal = {Phys. Rev. Research},
  volume = {3},
  pages = {L032034},
  year = {2021},
  doi = {10.1103/PhysRevResearch.3.L032034}
}

@article{HiuraSasa2021,
  author = {Hiura, Ken and Sasa, Shin-ichi},
  title = {Kinetic uncertainty relation on first-passage time for accumulated current},
  journal = {Phys. Rev. E},
  volume = {103},
  pages = {L050103},
  year = {2021},
  doi = {10.1103/PhysRevE.103.L050103}
}

@article{KacserBurns1973,
  author = {H. Kacser and J. A. Burns},
  title = {The control of flux},
  journal = {Symp. Soc. Exp. Biol.},
  volume = {27},
  pages = {65--104},
  year = {1973}
}

@article{HeinrichRapoport1974,
  author = {R. Heinrich and T. A. Rapoport},
  title = {A linear steady-state treatment of enzymatic chains: General properties, control and effector strength},
  journal = {Eur. J. Biochem.},
  volume = {42},
  pages = {89--95},
  year = {1974},
  doi = {10.1111/j.1432-1033.1974.tb03318.x}
}

@article{Fell1992,
  author = {D. A. Fell},
  title = {Metabolic control analysis: A survey of its theoretical and experimental development},
  journal = {Biochem. J.},
  volume = {286},
  pages = {313--330},
  year = {1992},
  doi = {10.1042/bj2860313}
}

@article{MelendezHeviaEtAl1990,
  author = {E. Mel{\'e}ndez-Hevia and N. V. Torres and J. Sicilia and H. Kacser},
  title = {Control analysis of transition times in metabolic systems},
  journal = {Biochem. J.},
  volume = {265},
  pages = {195--202},
  year = {1990},
  doi = {10.1042/bj2650195}
}

@article{McKeithan1995,
  author = {T. W. McKeithan},
  title = {Kinetic proofreading in {T}-cell receptor signal transduction},
  journal = {Proc. Natl. Acad. Sci. U.S.A.},
  volume = {92},
  pages = {5042--5046},
  year = {1995},
  doi = {10.1073/pnas.92.11.5042}
}

@article{EslamiMossallam2022,
  author = {B. Eslami-Mossallam and M. Klein and C. {van der Smagt} and K. {van der Sanden} and Jones, Jr., S. K. and J. A. Hawkins and I. J. Finkelstein and M. Depken},
  title = {A kinetic model predicts {SpCas9} activity, improves off-target classification, and reveals the physical basis of targeting fidelity},
  journal = {Nat. Commun.},
  volume = {13},
  pages = {1367},
  year = {2022},
  doi = {10.1038/s41467-022-28994-2}
}

@article{GromadskiRodnina2004,
  author = {K. B. Gromadski and M. V. Rodnina},
  title = {Kinetic determinants of high-fidelity {tRNA} discrimination on the ribosome},
  journal = {Mol. Cell},
  volume = {13},
  pages = {191--200},
  year = {2004},
  doi = {10.1016/S1097-2765(04)00005-X}
}

@article{GingrichHorowitz2017,
  author = {T. R. Gingrich and J. M. Horowitz},
  title = {Fundamental Bounds on First Passage Time Fluctuations for Currents},
  journal = {Phys. Rev. Lett.},
  volume = {119},
  pages = {170601},
  year = {2017},
  doi = {10.1103/PhysRevLett.119.170601}
}

@article{OwenGingrichHorowitz2020,
  author = {J. A. Owen and T. R. Gingrich and J. M. Horowitz},
  title = {Universal Thermodynamic Bounds on Nonequilibrium Response with Biochemical Applications},
  journal = {Phys. Rev. X},
  volume = {10},
  pages = {011066},
  year = {2020},
  doi = {10.1103/PhysRevX.10.011066}
}

@article{SartoriPigolotti2013,
  author = {P. Sartori and S. Pigolotti},
  title = {Kinetic versus Energetic Discrimination in Biological Copying},
  journal = {Phys. Rev. Lett.},
  volume = {110},
  pages = {188101},
  year = {2013},
  doi = {10.1103/PhysRevLett.110.188101}
}

@article{RaoPeliti2015,
  author = {R. Rao and L. Peliti},
  title = {Thermodynamics of accuracy in kinetic proofreading: Dissipation and efficiency trade-offs},
  journal = {J. Stat. Mech.: Theory Exp.},
  volume = {2015},
  number = {6},
  pages = {P06001},
  year = {2015},
  doi = {10.1088/1742-5468/2015/06/P06001}
}

@article{KumarBanerjeeGangopadhyay2022,
  author = {P. Kumar and K. Banerjee and G. Gangopadhyay},
  title = {Interplay of energy, dissipation, and error in kinetic proofreading: Control via concentration and binding energy},
  journal = {Physica A},
  volume = {603},
  pages = {127735},
  year = {2022},
  doi = {10.1016/j.physa.2022.127735}
}

@article{AslyamovEsposito2024,
  author  = {Aslyamov, Timur and Esposito, Massimiliano},
  title   = {General Theory of Static Response for {Markov} Jump Processes},
  journal = {Phys. Rev. Lett.},
  volume  = {133},
  pages   = {107103},
  year    = {2024},
  doi     = {10.1103/PhysRevLett.133.107103},
}

@article{AslyamovPtaszynskiEsposito2025,
  author  = {Aslyamov, Timur and Ptaszy{\'n}ski, Krzysztof and Esposito, Massimiliano},
  title   = {Nonequilibrium Fluctuation-Response Relations: From Identities to Bounds},
  journal = {Phys. Rev. Lett.},
  volume  = {134},
  pages   = {157101},
  year    = {2025},
  doi     = {10.1103/PhysRevLett.134.157101},
}

@article{ZhengLu2025a,
  author  = {Zheng, Jiming and Lu, Zhiyue},
  title   = {Universal Response Inequalities Beyond Steady States via Trajectory Information Geometry},
  journal = {Phys. Rev. E},
  volume  = {112},
  pages   = {L012103},
  year    = {2025},
  doi     = {10.1103/scg2-qkxv},
}

@article{ZhengLu2025b,
  author  = {Zheng, Jiming and Lu, Zhiyue},
  title   = {Unified Linear Fluctuation-Response Theory Arbitrarily Far from Equilibrium},
  journal = {Phys. Rev. E},
  volume  = {112},
  pages   = {064103},
  year    = {2025},
  doi     = {10.1103/rgys-zxgf},
}

@article{KwonChunParkLee2025,
  author  = {Kwon, Euijoon and Chun, Hyun-Myung and Park, Hyunggyu and Lee, Jae Sung},
  title   = {Fluctuation-response inequalities for kinetic and entropic perturbations},
  journal = {Phys. Rev. Lett.},
  volume  = {135},
  pages   = {097101},
  year    = {2025},
  doi = {10.1103/h45s-4118}
}

@article{LiuGu2025,
  author  = {Liu, Kangqiao and Gu, Jie},
  title   = {Dynamical activity universally bounds precision of response in {Markovian} nonequilibrium systems},
  journal = {Commun. Phys.},
  volume  = {8},
  pages   = {62},
  year    = {2025},
  doi = {10.1038/s42005-025-01982-w}
}

@article{Dechant2026,
  author  = {Dechant, Andreas},
  title   = {Finite-Frequency Fluctuation-Response Inequality},
  journal = {Phys. Rev. Lett.},
  volume  = {136},
  pages   = {207101},
  year    = {2026},
  doi     = {10.1103/3hs9-dz3d},
}

@article{GaoChunHorowitz2024,
  author  = {Gao, Qi and Chun, Hyun-Myung and Horowitz, Jordan M.},
  title   = {Thermodynamic constraints on kinetic perturbations of homogeneous driven diffusions},
  journal = {Europhys. Lett.},
  volume  = {146},
  pages   = {31001},
  year    = {2024},
  doi     = {10.1209/0295-5075/ad40cd},
}

@misc{FancherHorowitz2026,
  author       = {Fancher, Sean and Horowitz, Jordan M.},
  title        = {Topological building blocks of nonequilibrium response},
  year         = {2026},
  eprint       = {2607.12096},
  archivePrefix= {arXiv},
  primaryClass = {cond-mat.stat-mech},
}

@article{KatayamaNagayamaIto2026,
  author  = {Katayama, Koya and Nagayama, Ryuna and Ito, Sosuke},
  title   = {Diagrammatic expressions for steady-state distribution and static responses in population dynamics},
  journal = {Phys. Rev. Research},
  volume  = {8},
  pages   = {013312},
  year    = {2026},
  doi     = {10.1103/fc35-47fs},
}

@misc{Liu2025MCA,
  author       = {Liu, Weijiu},
  title        = {Simple Proofs of the Summation and Connectivity Theorems in Metabolic Control Analysis},
  year         = {2025},
  eprint       = {2501.12519},
  archivePrefix= {arXiv},
  primaryClass = {q-bio.MN},
}

@misc{Sekimoto2021,
  author       = {Sekimoto, Ken},
  title        = {Derivation of the First Passage Time Distribution for {Markovian} Process on Discrete Network},
  year         = {2021},
  eprint       = {2110.02216},
  archivePrefix= {arXiv},
  primaryClass = {cond-mat.stat-mech},
}

@article{AslyamovEsposito2024a,
  author  = {Aslyamov, Timur and Esposito, Massimiliano},
  title   = {Nonequilibrium Response for {Markov} Jump Processes: Exact Results and Tight Bounds},
  journal = {Phys. Rev. Lett.},
  volume  = {132},
  pages   = {037101},
  year    = {2024},
  doi     = {10.1103/PhysRevLett.132.037101},
}

@article{GoerlichEtAl2026,
  author  = {Goerlich, R{\'e}mi and Tartar, Antoine and Roichman, Yael and Sokolov, Igor M.},
  title   = {Fluctuation-response relation for a nonequilibrium system with resolved {Markovian} embedding},
  journal = {Phys. Rev. E},
  volume  = {114},
  pages   = {014138},
  year    = {2026},
  doi     = {10.1103/kldw-kftz},
}

@article{KlettLindner2025,
  author  = {Klett, Kolja and Lindner, Benjamin},
  title   = {Fluctuation-response relations and response-response relations for membrane voltage and spike train of stochastic integrate-and-fire neurons},
  journal = {Phys. Rev. E},
  volume  = {112},
  pages   = {044404},
  year    = {2025},
  doi     = {10.1103/crxw-pvcj},
}

@article{WangWangRen2026,
  author  = {Wang, Zi and Wang, Chen and Ren, Jie},
  title   = {Sensitivity analysis of cycle flux response in nonequilibrium dynamics},
  journal = {J. Chem. Phys.},
  volume  = {165},
  number  = {4},
  pages   = {044105},
  year    = {2026},
  doi     = {10.1063/5.0339421},
}

@article{KhodabandehlouMaesNetocny2025,
  author  = {Khodabandehlou, Faezeh and Maes, Christian and Neto{\v c}n{\'y}, Karel},
  title   = {Affine relationships between steady currents},
  journal = {J. Phys. A: Math. Theor.},
  volume  = {58},
  pages   = {155002},
  year    = {2025},
  doi     = {10.1088/1751-8121/adc8ea},
}

@article{HarunariEtAl2024,
  author  = {Harunari, Pedro E. and {Dal Cengio}, Sara and Lecomte, Vivien and Polettini, Matteo},
  title   = {Mutual Linearity of Nonequilibrium Network Currents},
  journal = {Phys. Rev. Lett.},
  volume  = {133},
  pages   = {047401},
  year    = {2024},
  doi     = {10.1103/PhysRevLett.133.047401},
}

@article{BebonSpeck2026,
  author  = {Bebon, Robin and Speck, Thomas},
  title   = {Mutual Linearity Is a Generic Property of Steady-State {Markov} Networks},
  journal = {Phys. Rev. Lett.},
  volume  = {136},
  pages   = {137401},
  year    = {2026},
  doi     = {10.1103/jcm3-57d8},
}

@article{DalCengioEtAl2025,
  author  = {{Dal Cengio}, Sara and Harunari, Pedro E. and Lecomte, Vivien and Polettini, Matteo},
  title   = {Mutual multilinearity of nonequilibrium network currents},
  journal = {SciPost Phys.},
  volume  = {19},
  pages   = {111},
  year    = {2025},
  doi     = {10.21468/SciPostPhys.19.4.111},
}

@article{BusielloLiangPigolotti2026,
  author  = {Busiello, Daniel Maria and Liang, Shiling and Pigolotti, Simone},
  title   = {Non-equilibrium symmetry of cyclic first-passage times},
  journal = {New J. Phys.},
  volume  = {28},
  pages   = {064602},
  year    = {2026},
  doi     = {10.1088/1367-2630/ae6fde},
}

\end{document}